%% file: kirchhoff.tex
\documentclass{gretsi}[english]
\usepackage[utf8]{inputenc}   
\usepackage[english]{babel}    
\usepackage{times}			
\usepackage{bm}
\usepackage{enumitem}
\usepackage{tikz}

\usepackage{amsmath,amssymb,amsthm,dsfont}
\usepackage{mathtools,mathdots,mathrsfs}
\usepackage{amsfonts}
\usepackage{hyperref}
\usepackage{graphicx}

\usepackage{etoolbox}
\patchcmd{\thebibliography}{\section*{\refname}}{}{}{}

\input{macros.tex}

\begin{document}
\title{Estimating a graph's spectrum via random Kirchhoff forests}

\author{
  \auteur{Simon}{Barthelmé}{simon.barthelme@gipsa-lab.grenoble-inp.fr}{1}
  \auteur{Fabienne}{Castell}{fabienne.castell@univ-amu.fr}{2}
  \auteur{Alexandre}{Gaudillière}{alexandre.gaudilliere@math.cnrs.fr}{3}
  \auteur{Clothilde}{Melot}{clothilde.melot@univ-amu.fr}{2}
  \auteur{Matteo}{Quattropani}{matteo.quattropani@uniroma3.it}{4}
  \auteur{Nicolas}{Tremblay}{nicolas.tremblay@cnrs.fr}{1,5}
}

\affils{\affil{1}{CNRS, Univ Grenoble-Alpes, Grenoble-INP, GIPSA-lab, Grenoble, France}\vspace{-.14cm}
  \affil{2}{Aix-Marseille Univ, I2M, Marseille, France}\vspace{-.14cm}
  \affil{3}{CNRS \& Aix-Marseille Univ, I2M, Marseille, France}\vspace{-.14cm}
  \affil{4}{Department of Mathematics \& Physics, Università di Roma Tre, Rome, Italy}\vspace{-.14cm}
  \affil{5}{Department of Mathematics \& Statistics, UiT the Arctic University of Norway, Troms\o, Norway}
}

\resume{\small{Pour des matrices de grande taille, la décomposition spectrale exacte est trop coûteuse, et on ne peut calculer toutes les valeurs propres. Un objectif plus modeste consiste à approcher la loi empirique des
  valeurs propres, pour connaître approximativement leur répartition. Les
  approches classiques utilisent des estimateurs de Monte Carlo
  pour obtenir une estimation des moments de la loi des valeurs propres. Dans
  cet article, nous introduisons une nouvelle approche pour ce problème, basée
  sur des forêts aléatoires sur graphes dites forêts de Kirchhoff. Nous montrons
  comment certaines observables de ces forêts peuvent être exploitées pour obtenir des estimateurs de la densité spectrale empirique de laplaciens de grands graphes. Si la précision souhait\'ee de l'estimation n'est pas trop importante, notre approche ouvre la voie à des estimations spectrales de graphes en temps sous-lin\'eaire en le nombre de liens.
    }}

\abstract{\small{Exact eigendecomposition of large matrices is very expensive, and it is practically impossible to compute exact eigenvalues.
  Instead, one may set a more modest goal of approaching the empirical
  distribution of the eigenvalues, recovering the overall shape of the
  eigenspectrum. Current approaches to spectral estimation typically work with
  \emph{moments} of the spectral distribution. These moments are first estimated
  using Monte Carlo trace estimators, then the estimates are combined to
  approximate the spectral density. In this article we show how
  \emph{Kirchhoff forests}, which are random forests on graphs, can be used to
  estimate certain non-linear moments of very large graph Laplacians. We show how to combine these moments into an
  estimate of the spectral density. If the estimate's desired precision isn't too high, our approach paves the way to the estimation of a graph's spectrum in time sublinear in the number of links. 
   }}

\maketitle
\sloppy

\small


\section{Introduction}
Computing {\let\thefootnote\relax\footnotetext{\hspace{-.6cm} Work partially funded by the ANR project GRANOLA (ANR-21-CE48-0009) and the European Union (Marie Curie project Rand4TrustPool, 101148828).}} the exact eigenvalue decomposition of a large symmetric matrix $\bL
\in \R^{n \times n}$ has cost $\O(n^3)$ in general, and becomes quickly out of
reach for matrices with $n > 10^4$. In some cases, all that is needed is an
estimate of the $p$ largest or smallest eigenvalues, and this can be achieved
using a variety of Krylov subspace methods \cite{saad2011numerical}. In other cases,
what is required is an estimate of the distribution of \emph{all} eigenvalues,
an important quantity in various applications \cite{weisse2006kernel}.

Specifically, letting $\lambda_1 \dots \lambda_n$ stand for the eigenvalues of
$\bL$, the goal is to estimate:
\begin{equation}
  \label{eq:cdf-lambda}
  c(\tau) = \frac{1}{n} \sum_{i=1}^n \Ind(\lambda_i \leq \tau)
\end{equation}
for all $\tau$. Here $c(\tau)$ measures the fraction of eigenvalues that are
smaller than the threshold $\tau$.

In what follows, it is useful to think of the function $c(\tau)$ as the
cumulative density function of the empirical distribution of the eigenvalues,
i.e. the discrete measure
\begin{equation}
  \label{eq:emp-measure}
  \mu = \frac{1}{n} \sum_i \delta_{\lambda_i}.
\end{equation}
To estimate the c.d.f. $c(\tau)$, the standard approach is to first estimate
\emph{moments} of the empirical distribution (the \emph{estimation} step) then
to reconstruct the cumulative density function from the estimated moments (the
\emph{reconstruction} step). 

In the existing literature, the estimation step takes advantage of the fact that
moments of the empirical measure $\mu$ correspond to traces of powers of
$\bL$:
\begin{equation}
  \label{eq:expectation}
  \mu(\lambda^j) \eqdef \int ^j \lambda^j \dd \mu (x) = \frac{1}{n} \sum_i \lambda_i^j = \frac{1}{n} Tr(\bL^j)
\end{equation}
Computing $\Tr(\bL^j)$ is expensive in general, but stochastic trace estimators
\cite{weisse2006kernel,martinsson2020randomized} like the Hutchinson-Girard estimator can be used. For any random vector
$\bz$ with $\E(\bz)=0$ and $\E(\bz \bz^t) = \bI$, we have:
\begin{equation}
  \label{eq:trace-estimatEion}
  \E(\bz^t \bL^j \bz) = Tr(\bL^j)
\end{equation}
and in this way traces can be computed by taking successive matrix-vector
products starting from a random vector $\bz$.

In the reconstruction step, one tries to reconstruct the cdf of $\mu$ from
knowledge of its $p$ first moments. Various techniques are available, the most effective are Jackson-Chebyshev expansions~\cite{di2016efficient} and the stochastic Lanczos quadrature~\cite{chen_analysis_2021}. The cost of these
techniques scales with the cost of a matrix-vector product $\bL \bz$, which
itself scales with $|\mathcal{E}|$, the number of edges of the graph. For the very sparse systems
that arise from discretising Partial Differential Equations, they are very
effective, but in denser systems they can be quite costly.

In what follows we describe a graph-theoretic approach for this problem that is
based on a specific random process called the ``Kirchhoff forest''. Kirchhoff
forests are a certain kind of random forests on graphs, with a range of
applications in randomised linear algebra and graph signal processing~\cite{pilavci2021graph, jaquard2024random, avena2020intertwining}. We
follow the same idea of estimating expectations under the empirical
distribution $\mu$, followed by a reconstruction step. However, our method is conceptually different from the literature: instead on relying on the computation of quadratic forms $\bz^t \bL^j \bz$ which prevents existing methods to improve the $|\mathcal{E}|$ term in the total computation cost, our method is based on sampling random forests of the graph and, under certain
assumptions, scales in $\O(n)$ rather than $|\mathcal{E}|$. The downside of this
is that, at this stage, our method is only competitive if one only needs a
medium-accuracy estimation of the graph's spectrum.

Section \ref{sec:kirchhoff-forests} introduces Kirchoff forests and how they can
be used to estimate certain spectral expectations. Section~\ref{sec:reconstruction} outlines the
reconstruction method. Section~\ref{sec:expes} gives some numerical results illustrating how our method outperforms the state-of-the-art if one only requires medium accuracy.

\section{Forests for moment estimation}
\label{sec:kirchhoff-forests}

In what follows we assume that $\bL$ is the Laplacian matrix of a weighted, undirected graph $G =
(\Ve, \Ed)$ with $n$ nodes and $|\mathcal{E}|$ edges:
\begin{equation}
  \label{eq:laplacian}
\bL = \bD - \bA
\end{equation}
where $\bA$ is the adjacency matrix and $\bD$ the degree matrix. See \cite{barthelmeinvtracest} for
how to extend this to diagonally-dominant $\bL$. We denote by $\mu(f)$ 
the expectation of function $f$ under the empirical spectral measure:
\begin{equation}
  \label{eq:empirical-expectation}
  \mu(f) = \frac{1}{n} \sum_{i}f(\lambda_i)
\end{equation}

Kirchhoff forests are random spanning forests on graphs that are particularly
easy to sample, and enable efficient estimators of various quantities tied to
the graph Laplacian \cite{barthelmeinvtracest,pilavci2021graph,Jaquard2023}.

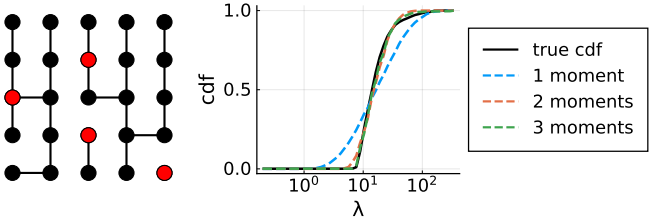
\begin{figure}
  \centering
  \begin{minipage}{.25\columnwidth}
  	\centering
   \begin{tikzpicture}[scale=0.5]
  	\foreach \x in {0,1,2,3,4} { \foreach \y in {0,1,2,3,4} {
  			\node[draw,circle,inner sep=2pt,fill=black] at (\x,\y) {}; } }
  	
  	\draw[thick, black] (2,2) -- (2,3) -- (2,4); \draw[thick, black] (2,2) --
  	(3,2) -- (3,1) -- (3,0); \draw[thick, black] (3,1) -- (4,1) -- (4,2) --
  	(4,3) -- (4,4); \draw[thick, black] (3,2) -- (3,3) -- (3,4);
  	
  	\draw[thick, black] (0,1) -- (0,2) -- (0,3) -- (0,4); \draw[thick, black]
  	(0,2) -- (1,2); \draw[thick, black] (0,0) -- (1,0) -- (1,1) -- (1,2) --
  	(1,3) -- (1,4);
  	
  	\draw[thick, black] (2,0) -- (2,1);
  	
  	\node[draw,circle,inner sep=2pt,fill=red] at (0,2) {};
  	\node[draw,circle,inner sep=2pt,fill=red] at (2,3) {};
  	\node[draw,circle,inner sep=2pt,fill=red] at (2,1) {};
  	\node[draw,circle,inner sep=2pt,fill=red] at (4,0) {};
  \end{tikzpicture}
  \end{minipage}
  \hfill
  \begin{minipage}{.7\columnwidth}
  	\centering~\\\vspace{0.4cm}
  	\includegraphics[width=\columnwidth]{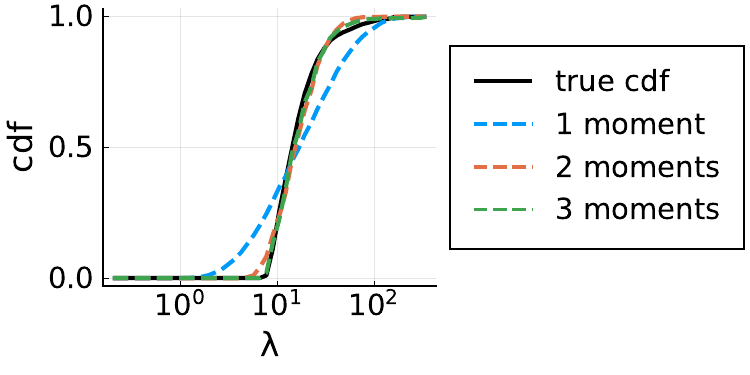}
  \end{minipage}
  \vspace{-0.2cm}
  \caption{\small{\textbf{(Left) Illustration of a rooted spanning forest} on the 5x5 grid graph, containing 
    4 trees, the smallest of which is a trivial tree of size 1. 
    Each tree contains a distinguished node called the root. Kirchhoff Forests (KFs) produce random rooted spanning forests with distribution given by eq. \eqref{eq:kirchhoff-forest}. \textbf{(Right) Illustration of a cdf estimation  of a graph's spectrum}, using our KF-based algorithm (for $l=1, 2, 3$ moments), on a Barabasi-Albert graph, with $n=10^3$ and  $\bar{d}=20$.}}\vspace{-0.2cm}
  \label{fig:fig1}
\end{figure}

In the context of graph theory, a \emph{spanning forest}
is a subgraph of $G$, made up of disjoint trees (acyclic subgraphs), and such
that every node belongs to the forest (the spanning part). The concept is
illustrated on fig. \ref{fig:fig1} (left).

A Kirchoff Forest (KF) is a distribution on \emph{rooted} spanning forests. A forest
is said to be rooted if every tree has a distinguished node called the root
(see fig. \ref{fig:fig1}). If the tree contains a single node, then the root of the tree is
the node itself. KFs are defined by the following probability mass
function:
\begin{equation}
  \label{eq:kirchhoff-forest}
  p_q(\phi) = \frac{|R(\phi)|^q }{Z_q} \prod_{ij \in \Ed(\phi)} w_{ij}
\end{equation}
Here $\phi$ is a forest with edge set $\Ed(\phi)$ and roots $R(\phi)$. $Z_q$ is
a normalisation constant, and $q$ is a parameter that governs the average number
of roots in the random forest. Large values of $q$ favor a large number of
roots. In fact, as shown in \cite{avena_two_2017}:
\begin{equation}
  \label{eq:mean-number-of-roots}
  \E( |R(\phi)| ) =  \sum_{i=1}^n \frac{q}{q+\lambda_i} = n \;\mu( q / (q + \cdot))
\end{equation}
\cite{barthelmeinvtracest} used this relationship to define inverse trace estimators. Here we go
further and show how to estimate various non-linear expectations under $\mu$. 
KFs have the advantage of being easy to sample. A basic way to sample a KF is
via an adaptation of Wilson's algorithm for Uniform Spanning Trees (see \cite{pilavci2021graph}). The
algorithm constructs trees branch-by-branch by successive aggregation of
loop-erased random walks. Wilson's algorithm for a KF with parameter $q$ runs in time $\O\left(\frac{|\mathcal{E}|}{q}\right)$. In our simulations, we use a more
sophisticated version, due to \cite{avena_two_2017}, the so-called Coupled Forests (CF) algorithm. The
CF algorithm produces a continuous sequence of random variables $\phi_q$, for $q \in
[q_{min},q_{max}]$, where each $\phi_q$ is a KF with parameter $q$. These random
variables are coupled (non-independent) but generating the whole family can be
done in time $\O\left(\frac{|\mathcal{E}|}{q_{min}}\right)$, the same as generating the most
expensive member. Unfortunately, the CF algorithm is too complex to be described
fully here. We refer the reader to \cite{avena_two_2017} and to our implementation in \emph{KirchhoffForests.jl}\footnote{available here: \url{github.com/dahtah/KirchhoffForests.jl}}. Sampling such a CF is the main building block of our proposed algorithm, and we will typically do so for $q_{min}$ set to a fraction of the average degree of the graph $q_{min} = \bar{d}/\alpha$, requiring a total $n$-linear computation cost of $\O\left(\alpha n\right)$.

Eq \eqref{eq:mean-number-of-roots} shows that certain observables of KFs
are spectral expectations of the form $\mu(f)$. Our idea is to estimate a large
set of such expectations, specifically,
\begin{equation}
  \label{eq:all-expectations}
  h(q,k) = \mu\left(  \left(\frac{q }{ q + \cdot}\right)^k \right)
\end{equation}
for $q \in [q_{min}, q_{max}]$, $k \in 1,2,\ldots,l$. In practice we use a
discrete grid of values of $q$ and $l$ is small (typically $l=3$ in our experiments). All these
expectations yield information about $\mu$, and can be used for reconstruction. 

To estimate these quantities, we use the following fact: define
\begin{equation}
  \label{eq:marginal-kernel}
  \bK_q = q ( q\bI + \bL)^{-1}; h(q,k) = \frac{1}{n} \Tr(\bK_q^k)
\end{equation}
and for a given KF $\phi_q$, the following matrix:
\begin{equation}
  \label{eq:est-marginal-kernel}
  \bM(\phi_q) = [ \Ind(r_{\phi_q}(i) = j)]_{i \in \Ve,j \in \Ve}
\end{equation}
where $r_{\phi_q}(i)$ is the root of node $i$ in the forest (the root of the
tree $i$ belongs to). Authors in 
\cite{avena_two_2017} have shown that
\begin{equation}
  \label{eq:unbiasedness}
  \E \left( \bM(\phi_q) \right) = \bK_q.
\end{equation}
In particular this implies eq. \eqref{eq:mean-number-of-roots}, which gives an
unbiased estimator for $h(q,1)$ (simply count the number of roots). To
generalise, we use the following lemma:
\begin{lemma}
  Let $\bX^{(1)}, \dots, \bX^{(k)}$ random i.i.d. matrices with expectation $\E(\bX) =
  \bM$. Then
  \begin{equation}
    \label{eq:eq-trace}
    \E\left( \Tr \left( \bX^{(1)} \dots \bX^{(k)} \right) \right) = \Tr\left( \bM^k \right)
  \end{equation}
\end{lemma}
\noindent \textit{Proof.}~\vspace{-0.5cm}
   \begin{align*}
 	\E&\left( \Tr \left( \bX^{(1)} \dots \bX^{(k)} \right) \right) = \E \left(\sum_{i=1}^n \sum_{j_1 \dots j_{k-1}} \bX^{(1)}_{i,j_1} \bX^{(2)}_{j_2,j_3} \dots \bX^{(k)}_{j_{k-1},i} \right) \\
 	&= 
 	\sum_{i=1}^n \sum_{j_1 \dots j_k} \bM(i,j_1)\dots \bM(j_{k-1},i) = Tr(\bM^k).
 \end{align*}
 ~\vspace{-1.5cm}\\
{\flushright{~\hfill\qedsymbol}}\\~\vspace{-0.06cm}

\noindent Applying this lemma to $\bM(\phi_q)$ defined in eq.
\eqref{eq:est-marginal-kernel}, we obtain:
\begin{theorem}
  Let $\phi_q^{(1)} \dots \phi_q^{(k)}$ be $k$ independent KFs with parameter $q$.
  Let
  $\rho_k = r_{\phi_q^{(k)}} \circ r_{\phi_q^{(k-1)}}  \dots \circ
  r_{\phi_q^{(1)}}$ the composition of the root maps, and
  $\th(q,k) = \frac{1}{n} \sum_{i=1}^n \Ind\left( \rho_k(i) = i \right) $. Then:
  \begin{equation}
    \label{eq:roots-of-roots}
    \E \left( \th(q,k)   \right)  =  \frac{1}{n} \Tr(\bK_q^k)  = h(q,k)
  \end{equation}
\end{theorem}
Given $k$ realisations of the KF $\phi_q$, $\th(q,k)$ is computable at cost $\O(kn)$. In addition, one can show that $\Var(\thqk) \leq h(q,k)/n$, thus ensuring fast convergence of the estimation. 

Our overall framework is as follows: pick a logarithmic grid for $q$, of size $n_\lambda$, ranging
from $q_{min}= \bar{d}/\alpha$ ($\alpha=100$ in our experiments) to $q_{max} = 2 d_{max}$. Recall that $\bar{d}$ (resp. $d_{max}$) refers to the
mean (resp. maximun) degree of the graph. Note that $2 d_{max}$ is a classical upper bound for $\lambda_{max}$,
obtained from applying the Gershgorin circle theorem to $\bL$. We run $l$ iterations of the Coupled Forests algorithm, to obtain a first estimate of $h(q,k)$ for $k$ in $1 \dots l$ at each value of $q$ in the grid. We replicate the whole 
process $s$ times to obtain several Monte-Carlo estimates that we finally average. At the end of the day, we obtain (very) precise estimates of $h(q,k)$ for all values of $q$ in the $n_\lambda$-sized grid and for $k\in[1\ldots l]$, for an overall computation cost of $\mathcal{O}\left((\alpha + n_\lambda) sl\;n \right)$. 
Several avenues are open to further improve the variance of these estimates, by adapting techniques from \cite{pilavci2021graph,pilavci2022variance} for instance. We now move on to
explaining how these estimates can be used for reconstruction of the spectral cdf $c(\tau)$.

\section{Reconstruction}
\label{sec:reconstruction}

At the end of the first phase, we have collected estimates of $h(q,k)$ for %
several values of $q$ and $k$. All of these contain some information about $\mu$
the spectral density of $\bL$, but extracting and combining that information is
not trivial. We will outline a procedure that we call ``fixed-q'' estimation,
which gives good results in practice. However, we do not wish to claim that it
is in any way optimal; the question of optimally combining the results of the
estimation phase is unsolved as yet. 
The starting point of the fixed-q procedure consists in noticing that if we
define:
\begin{equation}
  \label{eq:qmap}
  x_i = \frac{q}{q+\lambda_i}
\end{equation}
then:
\begin{equation}
  \label{eq:transformed-moments}
  h(q,k) = \frac{1}{n} \sum \left(\frac{q}{q+\lambda_i}\right)^k = \frac{1}{n} \sum x_i^k
\end{equation}
which are the classical moments of the empirical distribution of the $x_i$'s.
We note this measure
\begin{equation}
  \label{eq:transformed-emp-measure}
  \nu_q = \frac{1}{n} \sum_{i} \delta_{x_i}
\end{equation}
Since we have access to the moments of $\nu_q$, we can apply the standard
toolset of classical moment theory to form an approximation. In our numerical
simulations we rely on a maximum entropy estimator, defined below. This allows
us to form an estimator $\hat{\nu}_q$ based on the estimated moments $\th(q,1)
\ldots \th(q,l)$.

The map from $\lambda$ to $x$ given by \ref{eq:qmap} is nonlinear, and it
sends all the eigenvalues that are small (compared to $q$) to $x \approx 1$,
and all those that are very large to $x \approx 0$. In
theory, the map is invertible and we could recover an estimate of $\mu$ from
$\hat{\nu}_q$ by a change of variables. However, we only have access to a few
(estimated) moments of $\nu_q$ and the change of variables may be unstable. What
we do instead is simply form a pointwise estimate at the point where the map
\ref{eq:qmap} has highest slope (in magnitude): for $\lambda = q$. Noting that the cdf in $q$ is 
$c(q) = \mu([0,q]) = 1-\nu_q([0,\frac{1}{2}])$, we set
\[ \tilde{c}(q) = \tilde{\nu}_q\left(\left[0,\frac{1}{2}\right]\right).\]

\noindent We collect these estimates for all values of $q$ in the grid, yielding
an estimate of the overall cdf $c(\tau)$. An example is shown in fig.~\ref{fig:fig1} (right).

\textit{Maximum entropy estimator.} Let us now explain how $\nu_q$ can be
approximated from knowledge of its $k$ first moments. Our approach is inspired
by the maximum-entropy principle, a widely-used heuristic for problems of this
type \cite{mead1984maximum}. The maximum entropy principle suggests that when
all that is known about a certain distribution are a fixed number of moments,
then a reasonable estimate of the unknown distribution is the distribution with
maximum entropy that has compatible moments. Importantly, for fixed $q$, we have
a lower bound on $x_i$, $x_i \geq \frac{q}{q+\lambda_{max}} \geq
\frac{q}{q+2d_{max}}$ (see eq. \eqref{eq:qmap}). This implies that the distribution $\nu_q$ has support on
$[\frac{q}{q+2d_{max}},1]$, and we can use this information to constrain our
problem further.

The maximum-entropy distribution with
support on $[a,b]$ and moments $\vm = [m_1 \dots m_l]^t$ is given by:
\begin{equation}
  \label{eq:maxent}
  \underset{\nu \in \calP_{[a,b]}(\vm)}{\argmin} KL(\nu,\nu_0)
\end{equation}
where $KL(\nu,\nu_0)$ is the KL divergence with respect to the uniform measure
on $[a,b]$, and $\calP_{[a,b]}(\vm)$ is the set of probability measures on $[a,b]$ compatible
with the moment constraints (i.e. such that $\nu(x^j) = m_j$ for $j \in 1
\dots  l$). As is well-known (\cite{Cover2005}, ch. 12), the solution of eq. \eqref{eq:maxent} belongs to an
exponential family of the form:
\begin{equation}
  \label{eq:exp-family}
  q_{\bb}(x) = \exp( \bb^t \bv(x) - \psi(\bb) )
\end{equation}
where $\bv(x) = [x^i]_{i=1}^n $ and $\psi(\bb) = \log \int_{a}^b \exp \left( \bb^t
  \bv(x) \right) \dd x $. The optimal $\bb$ can be found by solving:
\begin{equation}
  \label{eq:moment-matching}
  \beta^{\star} = \underset{\bb \in \R^{s}}{\argmin\ } \psi(\bb)-\bb^t\vm
\end{equation}
Given the solution  $\bb_\star$, we can estimate  $\nu([a,\frac{1}{2}])$ as $
  \int_{a}^{\frac{1}{2}} \exp( \bb_\star^t \bv(x) - \psi(\bb_\star) ) \dd x$. 
  
  Now, there are a number of statistical and numerical issues to deal with. First of all, computing
$\psi(\bb)$ involves an intractable integral that we approximate by
Gauss-Legendre quadrature (\cite{golub2009matrices}, ch. 6). More importantly, the moments are only known
approximately. We define a confidence ellipsoid $B$ of the form:
\[ B = \left\{ \vect{m}' \in \R^l \text{ such that } \sum_{i=1}^l v_i^{-1}\left(m_i' - m_i\right)^2 \leq 1 \right\}\]
where $v_i$ is an upper-bound on the variance of the estimate $m_i$. When
solving eq. \ref{eq:moment-matching}, we interrupt the optimisation once the
moments corresponding to the iterate $\bb_t$ are in $B$. 
In addition, a vector of estimated moments $\vm$ can fall outside the set of
valid moments $\calM_l([a,b])$ \cite{Wu2020}. For instance, for all probability
distributions, we always have $\nu(x^2) \geq \nu(x)^2$, but estimated moments may
have $m_2 < m_1$ because of the noise. To deal with this, we proceed as follows:

1. We first test that the estimated moments are in $\calM_l([a,b])$. If they are, we proceed with the maximum entropy estimate. 

2. Otherwise, we project the estimated moments $\vm$ on $\calM_l([a,b])$
  (which is a form of denoising). We then apply maximum entropy to the denoised moments.\smallskip

Step 1 involves checking classical admissibility conditions for truncated moment
sequences, as described for instance in \cite{Schmuedgen2017}, p. 230. It
involves checking that two specific matrices formed from $\vm$ are
positive-definite. The denoising in Step 2 can be carried out via a
reformulation as a semidefinite program, too involved to be described here, see
\cite{Wu2020} or \cite{lasserre2008semidefinite}.

\section{Numerical experiments}
\label{sec:expes}

\begin{figure*}[t]
	\centering
	\begin{minipage}{.088\textwidth}
		\centering
		~\\
		$n=1000$\\\vspace{1.8cm}
		$n=5000$\\\vspace{1.8cm}
		$n=10000$
	\end{minipage}\hspace{0.4cm}
	\begin{minipage}{.16\textwidth}
		\centering
		2d Grid\\\medskip
		\includegraphics[width=\textwidth]{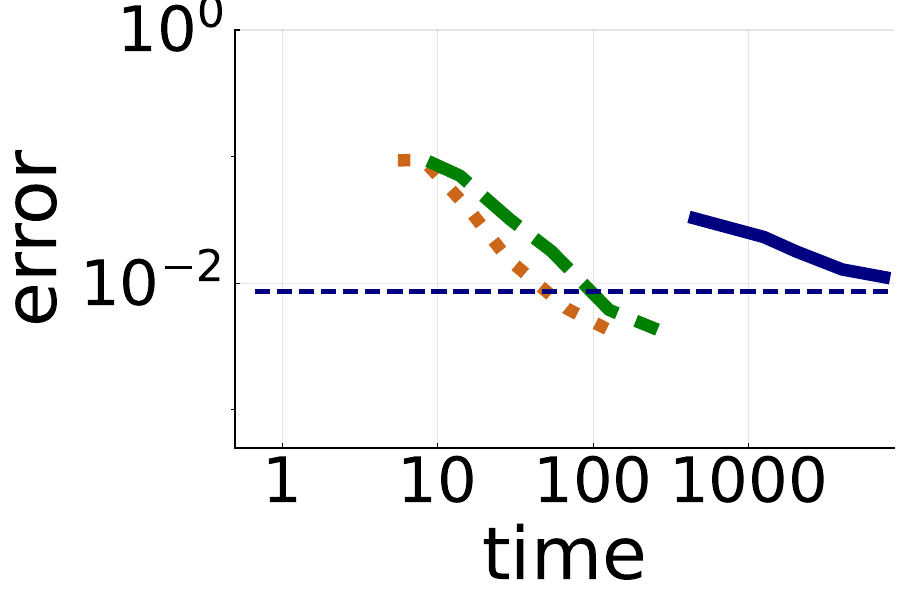}\\\smallskip
		\includegraphics[width=\textwidth]{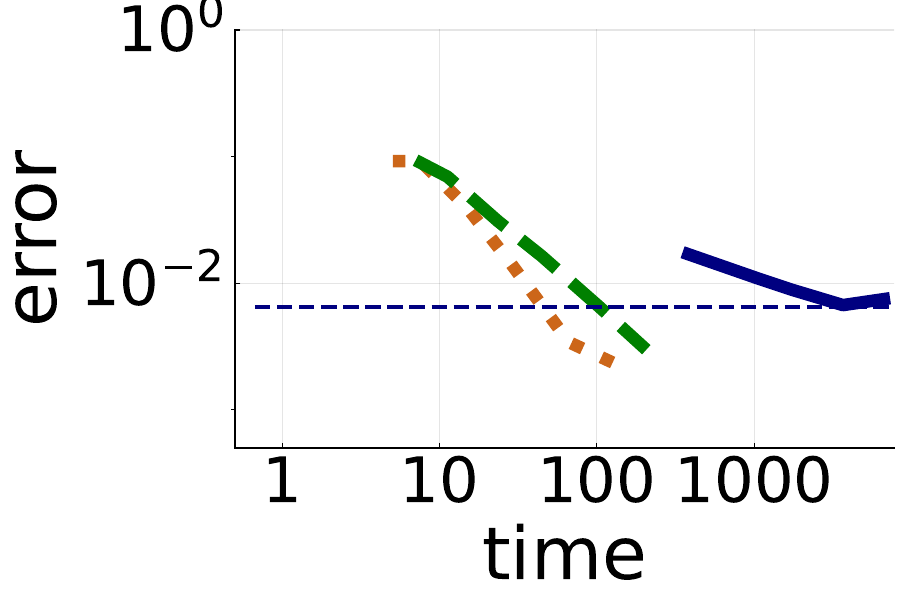}\\\smallskip
		\includegraphics[width=\textwidth]{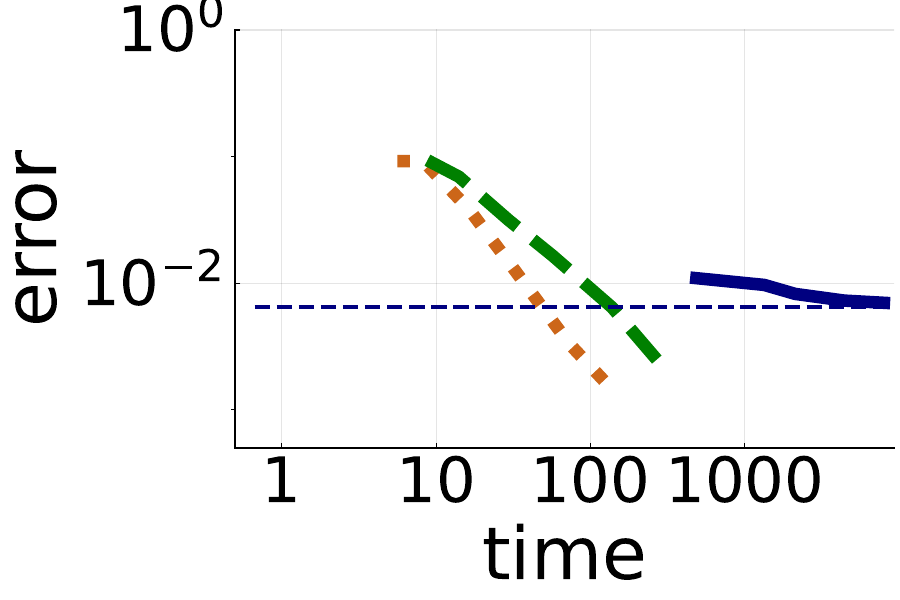}
	\end{minipage}\hspace{0.3cm}
	\begin{minipage}{.16\textwidth}
		\centering
		Sparse ER\\\medskip
		\includegraphics[width=\textwidth]{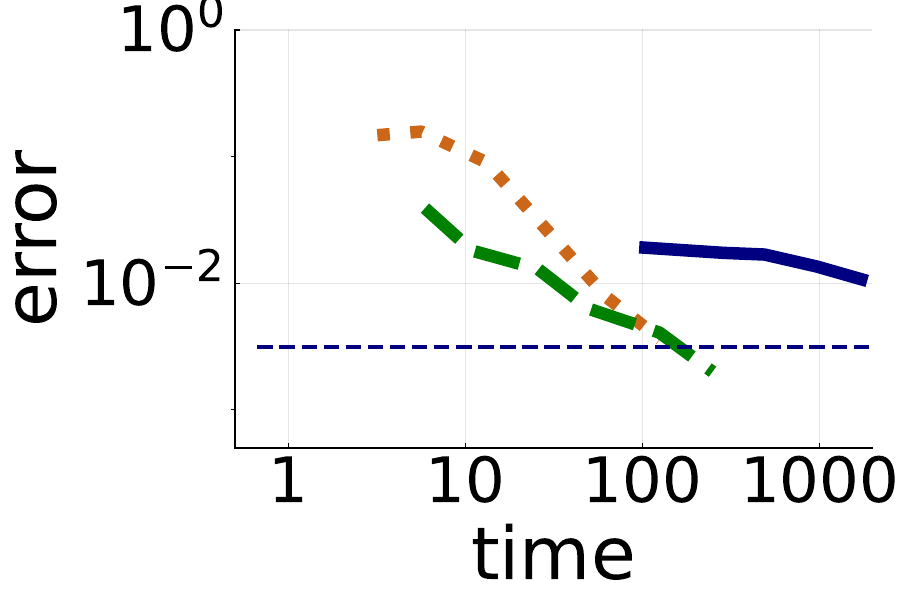}\\\smallskip
		\includegraphics[width=\textwidth]{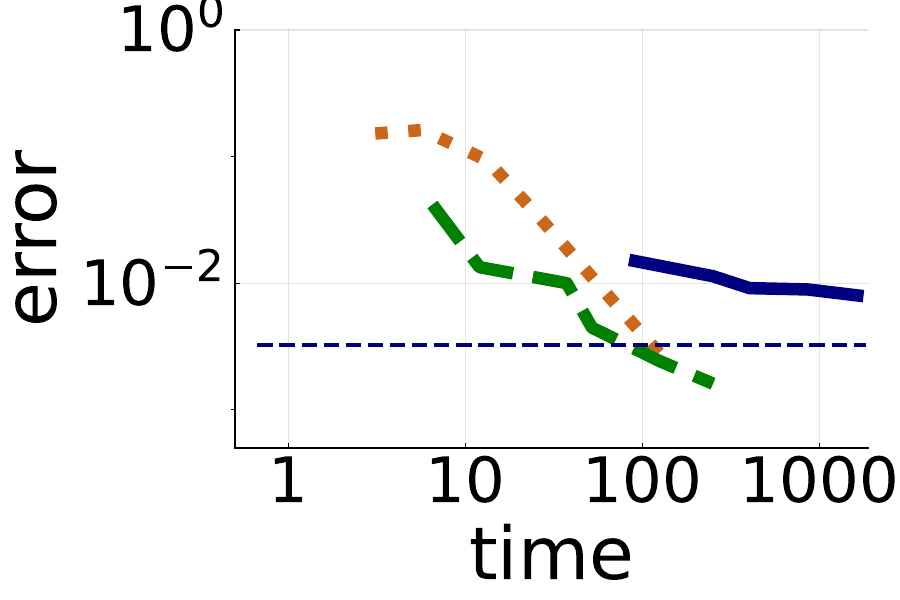}\\\smallskip
		\includegraphics[width=\textwidth]{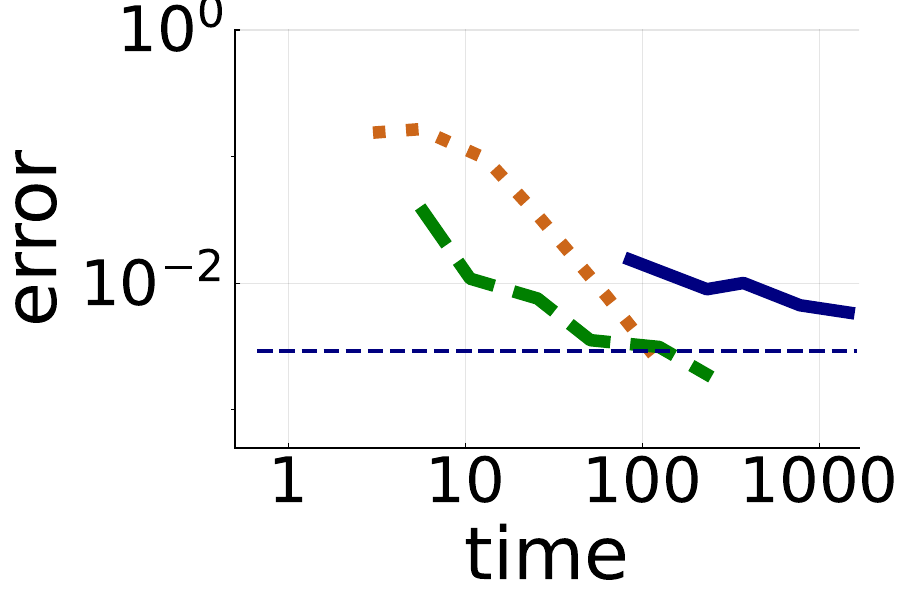}
	\end{minipage}\hspace{0.3cm}
	\begin{minipage}{.16\textwidth}
		\centering
		Sparse BA\\\medskip
		\includegraphics[width=\textwidth]{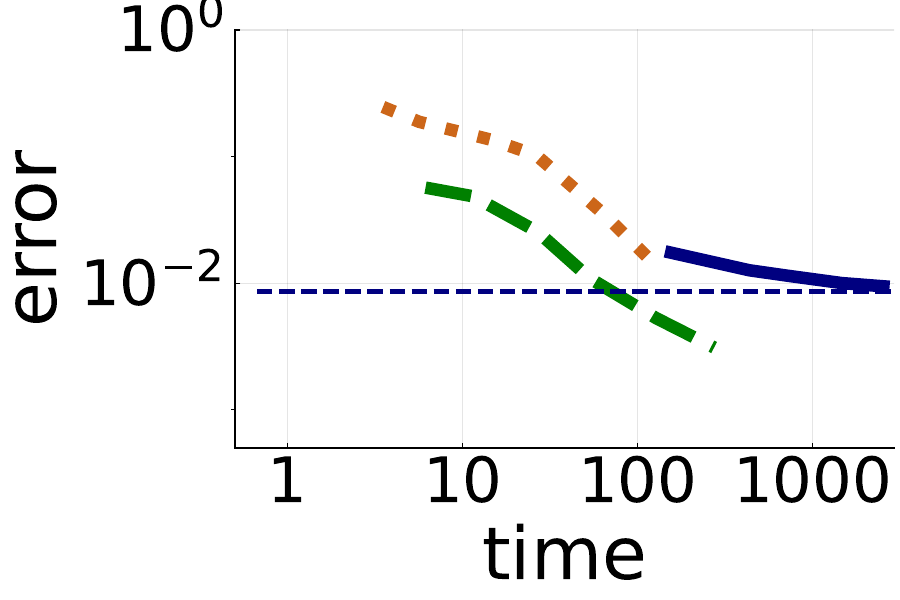}\\\smallskip
		\includegraphics[width=\textwidth]{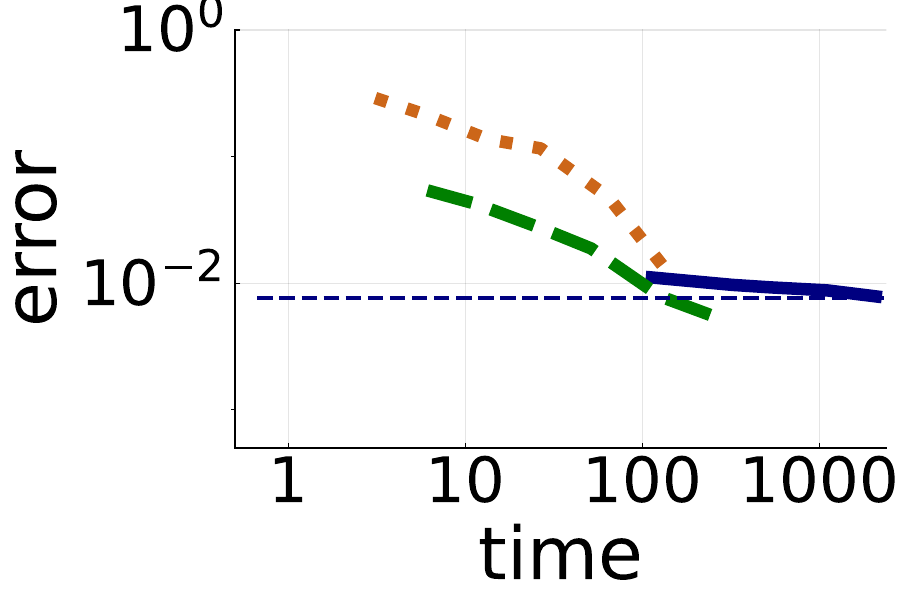}\\\smallskip
		\includegraphics[width=\textwidth]{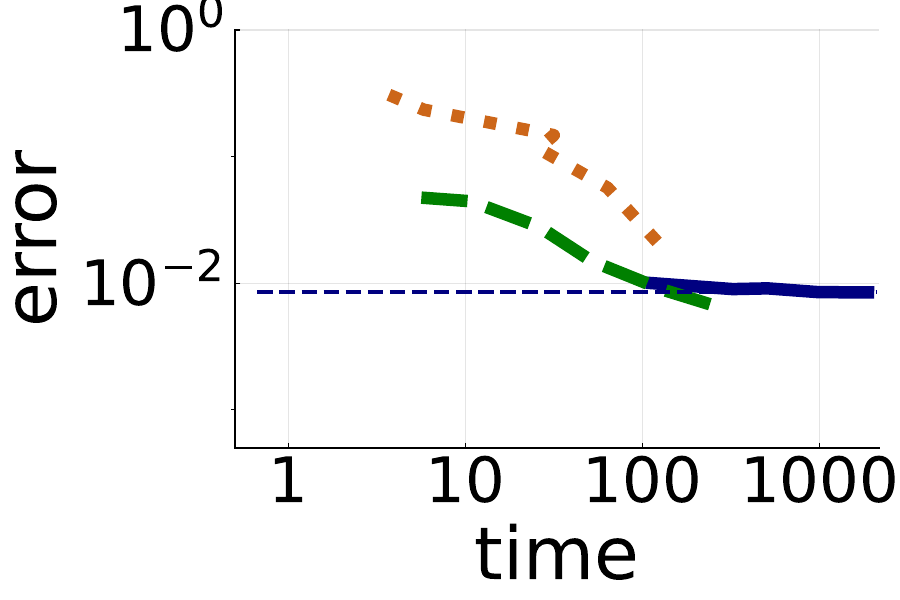}
	\end{minipage}\hspace{0.3cm}
	\begin{minipage}{.16\textwidth}
		\centering
		Dense ER\\\medskip
		\includegraphics[width=\textwidth]{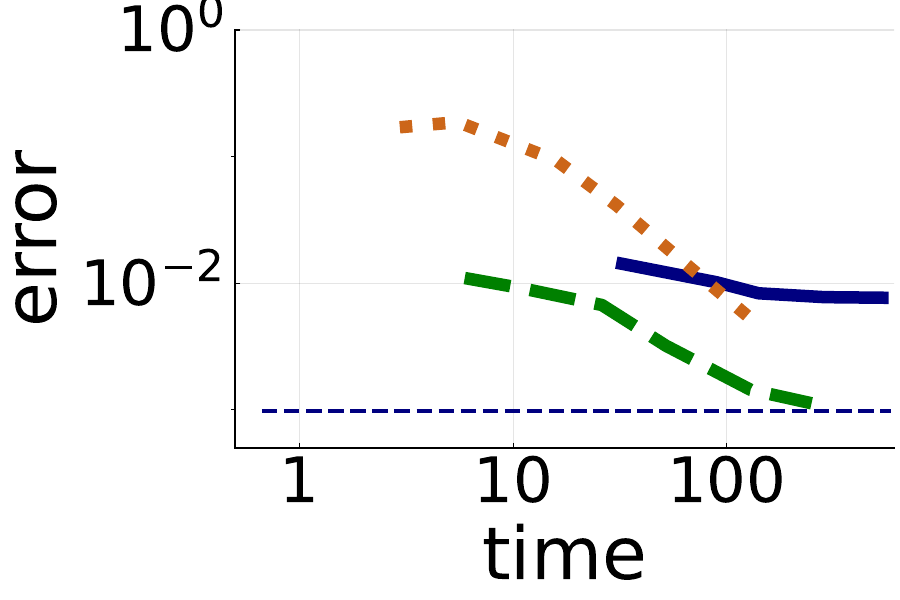}\\\smallskip
		\includegraphics[width=\textwidth]{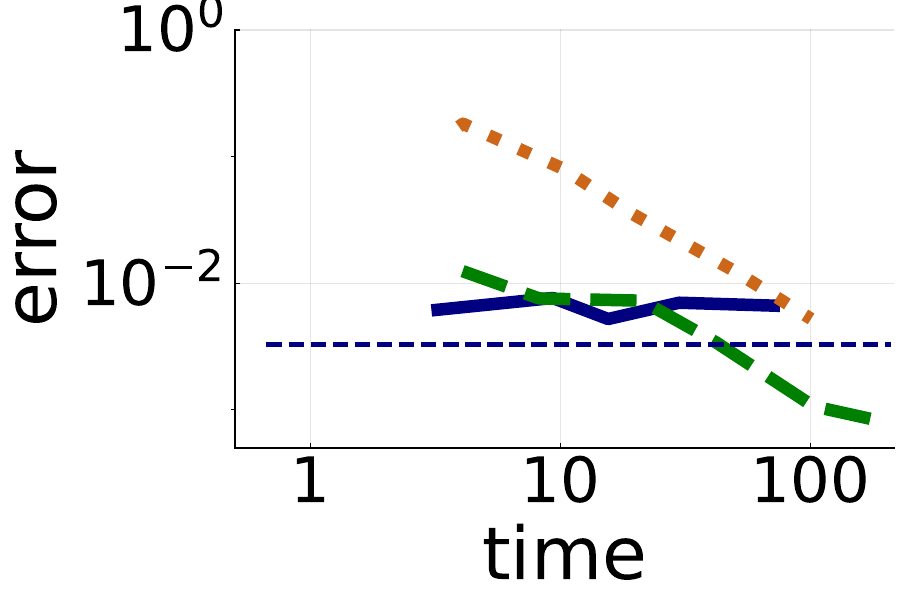}\\\smallskip
		\includegraphics[width=\textwidth]{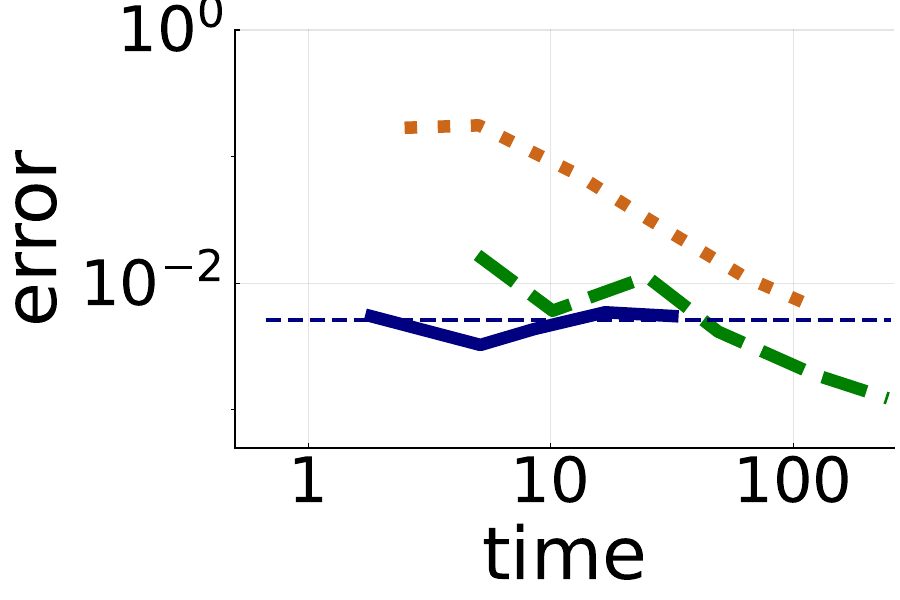}
	\end{minipage}\hspace{0.3cm}
	\begin{minipage}{.16\textwidth}
		\centering
		Dense BA\\\medskip
		\includegraphics[width=\textwidth]{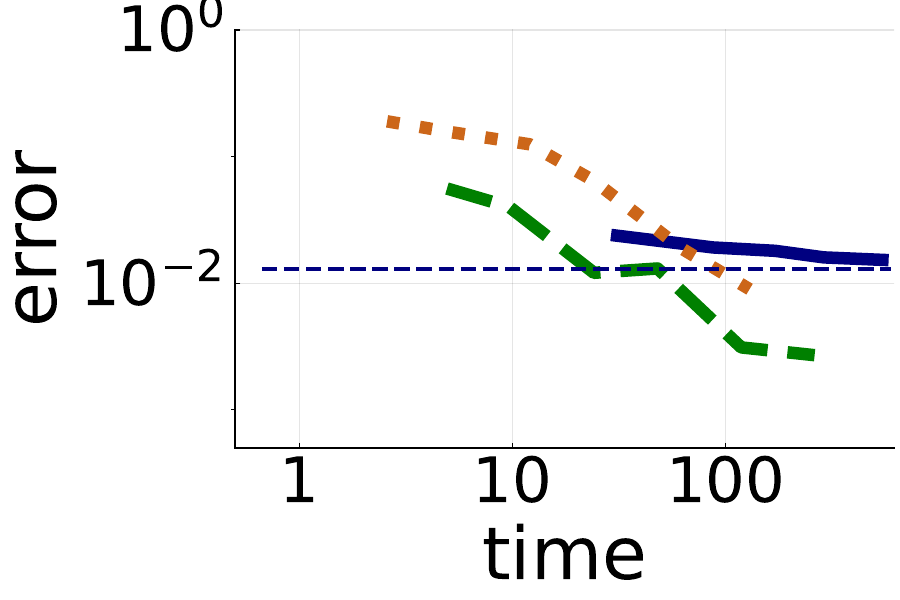}\\\smallskip
		\includegraphics[width=\textwidth]{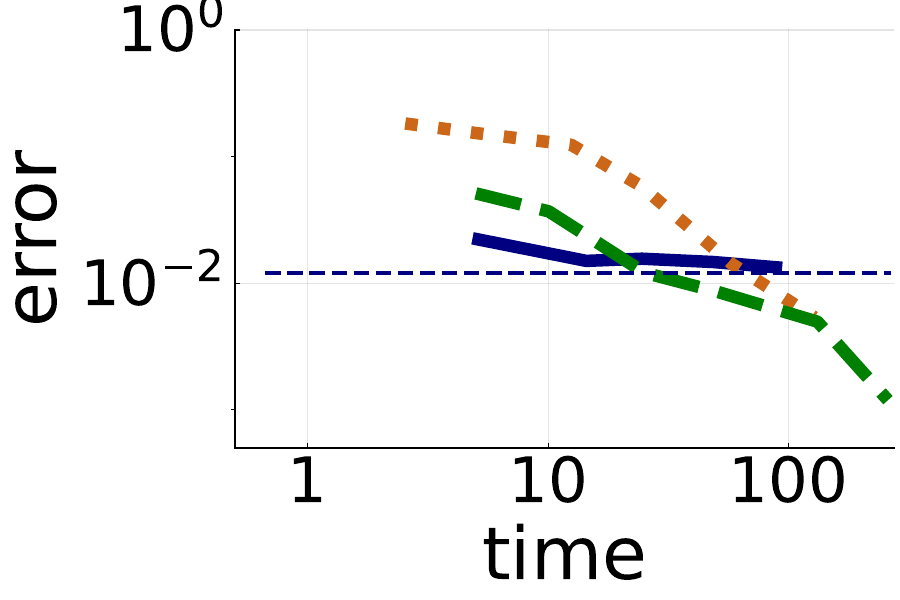}\\\smallskip
		\includegraphics[width=\textwidth]{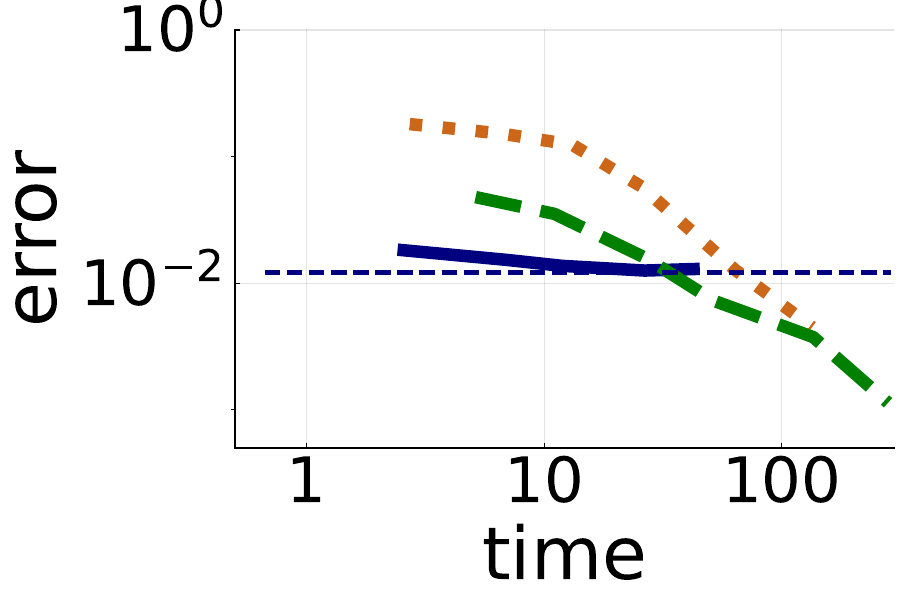}
	\end{minipage}
	\vspace{-0.25cm}
	\caption{\small{\textbf{Estimation error versus computation time} for 3 types of methods: \texttt{forests} (ours, in solid blue), \texttt{poly} (dotted orange), and \texttt{slq} (dashed green). The dotted horizontal blue line is the result of our reconstruction algorithm if we feed it the exact moments (rather than the KF-estimated ones). Each column is for a different type of graph: ER stands for Erdös-Renyi, BA for Barabasi-Albert, ``sparse'' means average degree of 20, and ``dense'' means average degree of $n/10$. Each line is for a different value of the number of nodes in the graph. For each graph and each value of $n$, the time axis is normalized by the time of computation of the corresponding matrix-vector multiplication $\bL \bx$. Results are averaged over $50$ realizations of all three methods (that are all stochastic), and $10$ realizations of each graph (for the 4 random graphs of the list).}}
	\vspace{-0.25cm}
	\label{fig:res2}
\end{figure*}

Fig.~\ref{fig:fig1} (right) illustrates a spectrum cdf estimation for a realization of a random Barabasi-Albert graph, with our method. We observe how the estimation improves as $l$ increases. Let us now compare our method to the state-of-the-art. The three methods we compare are:

$\bullet$ \texttt{poly}: a polynomial approximation method with Jackson-Chebychev damping coefficients, as in~\cite{di2016efficient}. \textit{Moment estimation step:} estimate the first $p$ moments of $L$ by averaging $\{\bx^t\bL^i\bx\}_{i\in[1..p]}$ over $r$ random vectors $\bx$ at an overall cost of $\mathcal{O}(rp|\mathcal{E}|)$ ($|\mathcal{E}|$ is equal to the number of non-zero entries of $\bL$). \textit{Reconstruction step:} for each value of the grid over $\lambda$ on which the cdf estimation is to be made, a precise linear combination of these estimates is computed, based on the Jackson-Chebychev approximation of the step function. 

$\bullet$ \texttt{slq}, short for Stochastic Lanczos Quadrature (see for instance~\cite{chen_analysis_2021}). \textit{Moment estimation\footnote{not truly a moment estimation step, but close enough to keep this term}  step:} for $r$ independent realizations of a random vector $\bx$, this method starts by creating a $p\times p$ tri-diagonal Lanczos matrix based on $\bx$ and its $p$ successive left-multiplication by $\bL$. This is done with an overall cost of $\mathcal{O}(rp|\mathcal{E}|)$. \textit{Reconstruction step:} these $r$ matrices are then diagonalized and results averaged. 

$\bullet$ \texttt{forests}: our method. \textit{Moment estimation step:} the moments $\{h(q,k)\}_{k=1..l}$ for $n_\lambda$ values of $q$ spaced between $\bar{d}/\alpha$ and $2d_\text{max}$ are estimated for an overall cost of $\mathcal{O}(\alpha ln)$ to sample the $l$ coupled forest + $\mathcal{O}(n_\lambda ln)$ for the computation of the $l$-th order roots at each $q$ of the grid of size $n_\lambda$ . This is repeated $s$ times to obtain Monte-Carlo averages, yielding an overall cost of $\mathcal{O}((\alpha+n_\lambda) sln)$. \textit{Reconstruction step:} as described  in Section~\ref{sec:reconstruction}.\smallskip

Note that the reconstruction cost of each method is independent of $n$, and thus
negligible asymptotically in $n$: in the following, we do not take into account
its computation cost as we want to show trends in large $n$. Also, note that in
the above, $s$ and $r$ are similar ``Monte-Carlo parameters'', and $p$ and $l$
are similar ``moment-order parameters''. For different sizes $n$ of graphs,
different types of graphs, and the parameter choices $\alpha=100$ and
$n_\lambda=15$, we plot in Fig.~\ref{fig:res2} the reconstruction error of the
cdf versus the computation time of each method. For \texttt{poly} and
\texttt{slq}, we show the results for $r=5$ and $p\in[1, 2, 5, 10, 25, 50]$
(each value of $p$ giving a data point in the error-computation time axes of the
figure) as we have observed in our examples that performance stabilizes quickly
in $r$. Regarding our method, it would seem sensible to fix the
Monte-Carlo parameter $s$ as above and make the moment parameter $l$ vary.
However, we are confronted with an important limitation of our method as it is
currently: whereas the moment estimation step could work with increasing $l$'s,
the reconstruction step starts to show signs of
instability\footnote{inherited from well-known instabilities of maximum-entropy
	algorithms} for $l\geq 4$. We thus only plot results for $l=3$ and
$s\in[1, 3, 5, 10, 20]$.

\begin{figure}
	\begin{minipage}{.42\columnwidth}
		\centering
		\includegraphics[width=\textwidth]{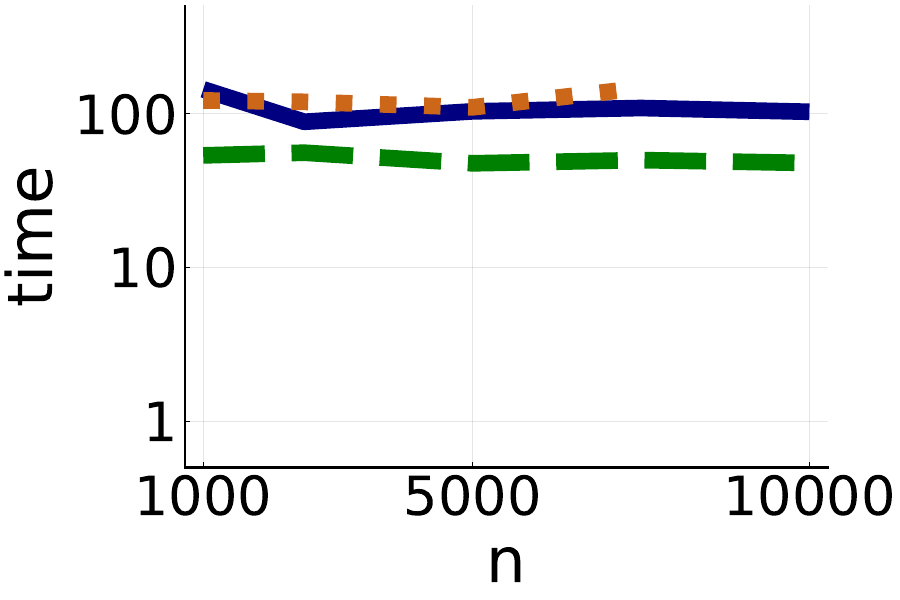}
	\end{minipage}\hfill
	\begin{minipage}{.57\columnwidth}
		\centering
		\includegraphics[width=\textwidth]{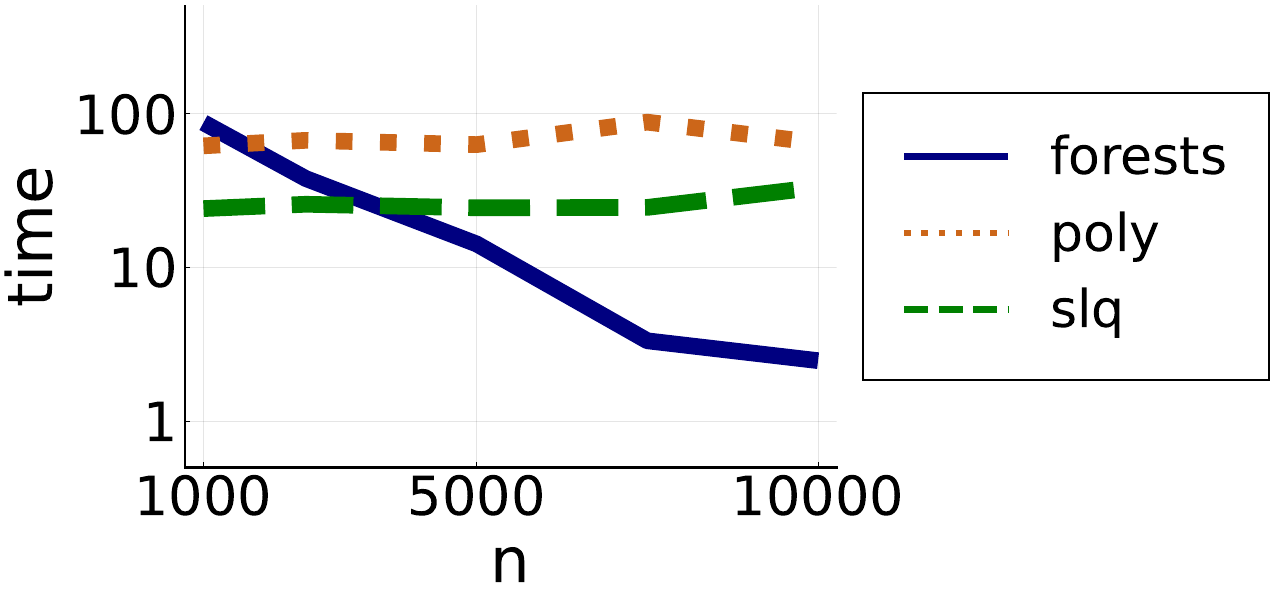}
	\end{minipage}
	\vspace{-0.2cm}
	\caption{\small{\textbf{Computation time required to reach a $2\%$  error versus the number of nodes}, for all three methods. Left: sparse BA graph. Right: dense BA graph. The time axis is normalized  by the time of computation of the  matrix-vector multiplication $\bL \bx$.}}
	\label{fig:res3}
	\vspace{-0.4cm}
\end{figure}

Even with this small, fixed value of $l$, we observe that if one accepts a
moderate precision in the estimation, then our method can provide significant
performance improvements. Note that this happens even though the error
reconstruction is necessarily lower bounded -- here by the horizontal blue lines
-- as opposed to the other two methods for which the error decreases
continuously with $p$. For instance, if one accepts a $2\%$ reconstruction
error, Fig.~\ref{fig:res3} shows the time needed to reach that error versus
the number of nodes for sparse and dense graphs. As expected from the previous theoretical time analyses of each method, the time to reach a given error is linear in the number of edges\footnote{The time to compute $\bL \bx$  is linear in $|\mathcal{E}|$} $|\mathcal{E}|$ for \texttt{poly} and \texttt{slq} and \textit{sublinear} in $|\mathcal{E}|$ (in fact, linear in $n$) for our method -- which is, we believe, remarkable.

\vspace{-0.35cm}
\section{Concluding remarks}\vspace{-0.2cm}
This is a first communication on a new and promising class of methods to estimate a graph's spectrum. Whereas state-of-the-art methods are classically based on successive matrix-vector multiplications $\bL\bx$ (and as such cannot pretend to go beyond a linear cost in $|\mathcal{E}|$), our results, by subtly subsampling the graph in several random spanning forests, pave the way to sublinear-time (in $|\mathcal{E}|$) spectrum estimation. We experimentally confirm this at least when moderate precision is enough, and when both the lowest $\lambda$ for which one wants an estimate of the cdf (in our experiments, we chose $\bar{d}/\alpha$ with $\alpha=100$) and the grid-size chosen for the spectrum's estimate, $n_\lambda$ (equal to $15$ in our experiments), are independent of $n$. 
Current work in progress is devoted to stabilizing the reconstruction algorithm and go beyond $l=3$ in order to increase our method's precision.

\vspace{-0.4cm}
\section*{References}
{
	\footnotesize{
		}}

\end{document}

%% file: macros.tex
\newcommand{\vect}[1]{\boldsymbol{#1}}
\newcommand{\matr}[1]{\boldsymbol{#1}}
\newcommand{\eqdef}{\stackrel{\textrm{def}}{=}}

\newcommand{\E}{\mathbb{E}} 
\newcommand{\Var}{\mathrm{Var}} 

\renewcommand{\th}{\tilde{h}}
\newcommand{\thqk}{\th(q,k)}

\newcommand{\bL}{\matr{L}}
  
  \newcommand{\bA}{\matr{A}}
  \newcommand{\bX}{\matr{X}}
  \newcommand{\bD}{\matr{D}}

  \newcommand{\bK}{\matr{K}}

  \newcommand{\bM}{\matr{M}}
  \newcommand{\bI}{\matr{I}}
  \newcommand{\bx}{\matr{x}}

    \newcommand{\Ve}{\mathcal{V}}
    \newcommand{\Ed}{\mathcal{E}}

    \newcommand{\R}{\mathbb{R}}

\renewcommand{\O}{\mathcal{O}}

\newcommand{\calM}{\mathcal{M}}
\newcommand{\calP}{\mathcal{P}}

\newcommand{\bb}{\vect{\beta}}
\newcommand{\bv}{\vect{v}}
\newcommand{\vm}{\vect{m}}

\DeclareMathOperator*{\Tr}{Tr}
\DeclareMathOperator*{\argmin}{\mathrm{argmin}}
\newcommand{\Ind}{\vect{\mathds{I}}}

\newcommand{\dd}{\text{d}}
\newcommand{\bz}{\vect{z}}

\newtheorem{definition}{Definition}[section]
\newtheorem{theorem}[definition]{Theorem}

\newtheorem{lemma}[definition]{Lemma}